\documentclass[nocompress]{spie}  

\usepackage{amsmath,amsfonts,amssymb}
\usepackage{graphicx}
\usepackage[colorlinks=true, allcolors=blue]{hyperref}

\title{EIFIS: a modular extreme integral field spectrograph\\for the 10.4m GTC}

\author[a]{Christina C. Th\"one}
\author[b]{Antonio de Ugarte Postigo}
\author[c]{Marisa Garc\'ia Vargas}
\author[d]{Jos\'e Feliciano Ag\"u\'i Fern\'andez}
\author[c]{Ana P\'erez Calpena}
\author[c]{Ernesto S\'anchez Blanco}
\author[c]{Manuel~Maldonado}

\affil[a]{Astronomical Institute of the Czech Academy of Sciences
(ASU-CAS), Fri\v{c}ova 298, Ond\v{r}ejov, 251 65, Czech Republic}
\affil[b]{Artemis, Observatoire de la C\^ote d'Azur, Universit\'e C\^ote d'Azur, Boulevard de l'Observatoire, 06304 Nice, France}
\affil[c]{FRACTAL, S.L.N.E., Spain}
\affil[d]{IAA-CSIC, Glorieta de la Astronom\'ia, s/n, 18008 Granada, Spain}

\authorinfo{Further author information: (Send correspondence to C.C.T.)\\C.C.T.: E-mail: christina.thoene@asu.cas.cz, \\  A.d.U.P.: E-mail: deugarte@oca.eu}

\begin{document} 
\maketitle

\begin{abstract}
EIFIS ({\it E}xtreme {\it I}ntegral {\it FI}eld {\it S}pectrograph) is a modular integral field spectrograph, based on image slicers, and makes use of new, large format detectors. The concept is thought to cover the largest possible field of view while producing spectroscopy over the complete optical range (3 000 - 10 000 {\AA}) at a medium resolving power of $\sim$2400. In the optimal concept, each module covers a field of view of 38" x 38" with 0.3" spaxels, which is fed into a double spectrograph with common collimator optics. The blue arm covers the spectral range between 3000 and 5600 Å and the red arm between 5400 and 10100 {\AA}, allowing for an overlap range. The spectra are imaged onto 9.2k x 9.2k detectors using a double pseudoslit. The proposed design for the 10.4m Gran Telescopio Canarias uses a total of 6 such modules to cover a total of 2.43 square arcminutes. Here we will present the conceptual design of the instrument and a feasibility study of the optical and mechanical design of the spectrographs. We discuss the limitations and alternative designs and its potential to produce leading edge science in the era of extremely large telescopes and the James Webb Space Telescope. 
\end{abstract}

\keywords{Integral field spectrograph, large FOV, feasibility study}

\section{INTRODUCTION}
\label{sec:intro}  
Integral field spectrographs (IFSs) deliver spectral information not only for a single object but in every point (called a ``spaxel'') of a two-dimensional area of the sky. They have become one of the workhorse instruments of modern astronomy such that almost any major telescope is equipped with an IFS of some kind. There are several commonly used designs for such spectrographs: image slicers that deliver rows of long-slit spectra spaced exactly by the slit width, which are then subdivided into one-dimensional spectra in each spaxel, examples are MUSE at the VLT \cite{Bacon2010} or WiFES at the 2.3m Siding Spring observatory \cite{Dopita07}. Alternatively, IFUs use microlenses or fibre-bundles to directly obtain a 1D spectrum in each spaxel after passing through dispersing optics, examples are MEGARA at the 10.4m GTC \cite{MEGARA} or the now decommissioned VIMOS spectrograph at the VLT \cite{LeFevre03}. For reasons of efficiency, the most widely used concept currently are image slicers.

The two main technical issues with this kind of instruments are the complexity of the optics to obtain spectra in every spaxel and, connected to that, the limitations due to the detector size needed to fit the spectra. For this reason, most IFSs only have a relatively small field-of-view (FOV) of a few square arcseconds, or large spaxel sizes and often a limited wavelength range. However, there are many science cases that would profit from a large FOV, from very extended objects such as nearby galaxies or objects and regions in the Milky Way to galaxy clusters and large, deep surveys to search and characterise high-redshift objects. There currently exist only very few such IFSs. The best example is MUSE at the VLT, which has been one of the most requested instruments at VLT since its installation. Its huge success has lead to the construction of a new IFS at VLT to extend the capabilities of MUSE, called ``BlueMUSE'' \cite{Richard2019}, to cover wavelengths down to 3500 \AA{} and a larger FOV. Larger FOV IFUs such as VIRUS exist but do not have full spatial coverage, e.g. VIRUS at the HET \cite{HillVIRUS} only has a filling factor of 1/9 of the field in one exposure and serves different science goals mainly targeted to long-term surveys of the entire sky. 

Here we present an instrument concept for an IFS combining a large FOV with a seeing adapted spaxel size and full coverage over the entire visible wavelength range, called ``EIFIS'' ({\it E}xtreme {\it I}ntegral {\it FI}eld {\it S}pectrograph). In addition to possible science applications we detail our technical feasibility study regarding optics and mechanics exploring different design concepts and required trade-offs to meet most of the high level requirements.

\section{SCIENCE GOALS}
The driving idea for this type of very large IFS is to observe stellar populations and even individual massive stars in galaxies out to several tens of Mpc while covering the entire galaxy in one or very few pointings. Many strong emission lines from warm gas excited by young massive stars are observable at visible wavelength, allowing to infer many properties of the gas such as abundances, star formation rate or extinction. Older stars contribute to the stellar continuum and absorption lines and allow to study the composition of the stellar population. MUSE, even in combination with the future BlueMUSE instrument, would usually require several pointings and observations with two different instruments (MUSE only covers the wavelength range $\sim$4800-9300\AA{}) to observe these galaxies, adding the complications of different observing conditions for different exposures. In addition, no such instrument is currently available in the Northern Hemisphere, excluding a number of very interesting objects. To achieve similar observations but with much lower efficiency and lower quality is to apply an old observing technique called ``drift scan'' or ``stepped slit'' spectroscopy recently performed for the TYPHOON survey at the 2.5m du Pont telescope on Las Campanas \cite{Ho2017}.

      \begin{figure} [!ht]
   \begin{center}
   \includegraphics[width=12cm]{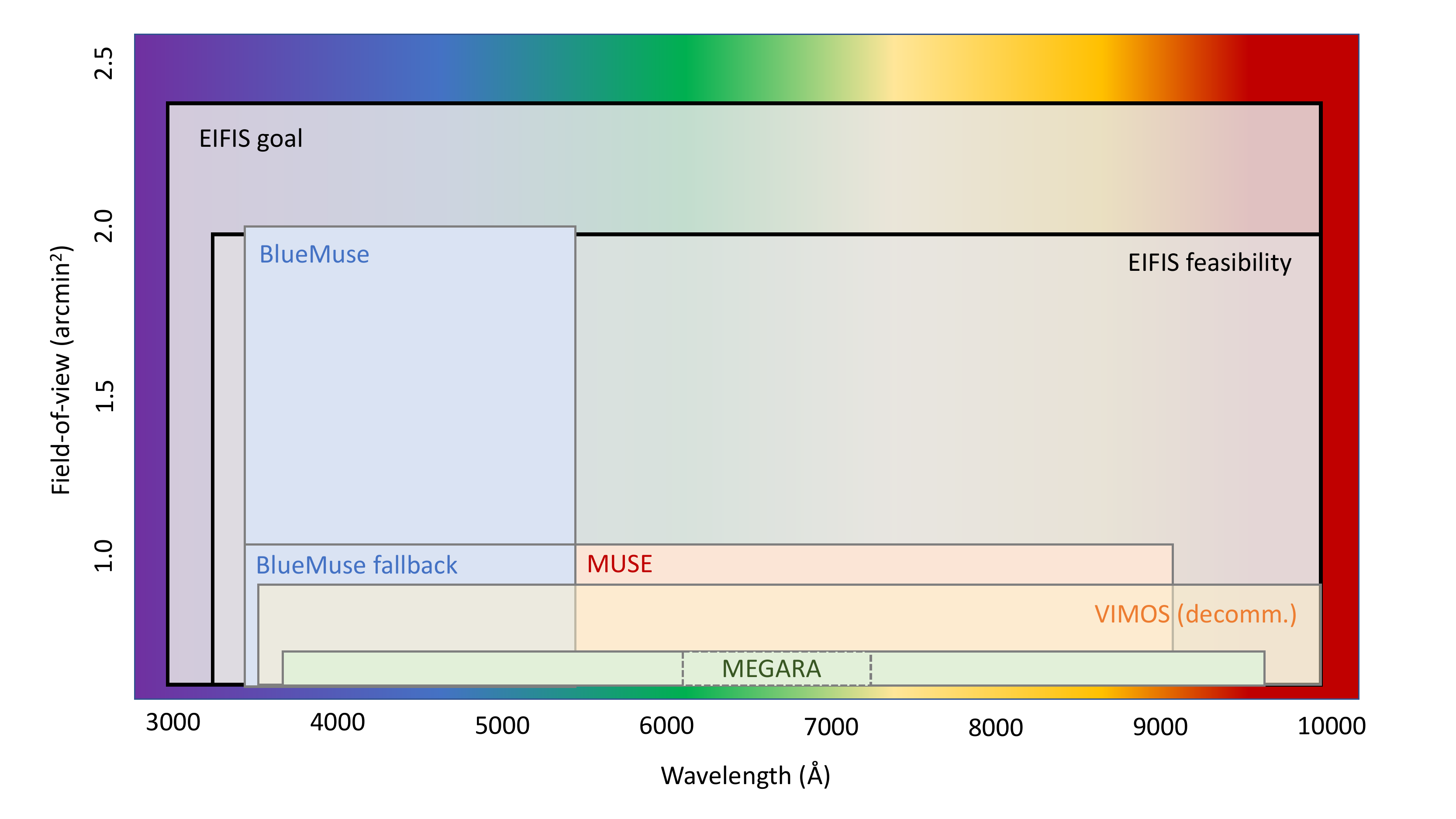}
	\end{center}
   \caption[] 
   { \label{fig:IFUs} 
Comparison of field of view and wavelength coverage for past, current and planned IFSs in 8-10 m class telescopes. In the cases of IFSs with multiple modes (MEGARA, VIMOS), we display a setup corresponding to the largest field of view and broadest wavelength coverage.}
   \end{figure} 

However, a large IFU with a FOV of 2-3 arcmin$^2$ can serve many other science topics in different astronomical communities. It would allow to study the gas large star-forming regions and stars in star clusters in the Milky Way and the Local group. It can also observe many objects in galaxy clusters simultaneously and study the large-scale structure of the Universe by determining redshifts of many galaxies in one observation. It would allow to detect and characterize transient sources with arcmin error boxes from triggers on high-energy satellites to neutrino observatories. A large-FOV IFU would be complementary to the instrumentation proposed for future 30m-class telescopes, which aims at much higher spatial resolution (in connection with adaptive optics) at the cost of FOV. The science of an IFU in the VIS range is also complementary to other efforts in the NIR both on ground (30m-class telescopes) and in space (JWST) regarding redshift coverage and properties that can be studied in either the VIS or NIR range. The scientific goals of EIFIS are still being developed and we explicitly welcome suggestions regarding further science topics as well as synergies with other facilities.

\section{EIFIS INSTRUMENT CONCEPT}

The main purpose of EIFIS is to observe a large FOV at full coverage in the entire visible range while retaining a moderate spatial and spectral resolution. To do this, we propose a modular design with multiple spectrographs located on the Nasmyth platform. The light from the telescope would go through a derotator that would ensure that the field is kept stable at the desired sky position angle. The light would then go through the image slicers that would then feed multiple spectrographs, that would work in parallel to fill a large sky area. The choice of image slicers allows to maximise efficiency, field of view and spatial resolution. Each of the spectrographs will be double, with a blue and a red arm that together would cover the full optical spectral range.

To accommodate the necessary number of pixels for such an instrument we aim at using the CCD290-99, which provides an area of 9.2$\times$9.2 kpixel at a pixel size of 10 $\mu$m. The spatial resolution of the spaxels in the spectral direction is limited by the width of the slitlets of the image slicer, whereas they are seeing limited in the spatial direction. As opposed to typical image slicers that re-image the area of the sky into a single pseudo-slit at the entrance of the spectrograph, EIFIS generates two parallel pseudo-slits to be imaged on the left and right half of the same detector. This doubles the area covered within a single spectrograph, but also limits the amount of pixels available to disperse the spectrum. However, given the area of the detectors, we can still disperse the spectral range over 4.5 kpixel, which is larger than other similar spectrographs such as MUSE with a single pseudo-slit. Allowing for a reasonable spacing between the different reprojected slits, this design could, in principle, reproject an area of the sky of 38" x 38" with 0.3" spaxels, before considering any further optomechanical limitations.

   \begin{figure} [!ht]
   \begin{center}
   \includegraphics[width=10cm]{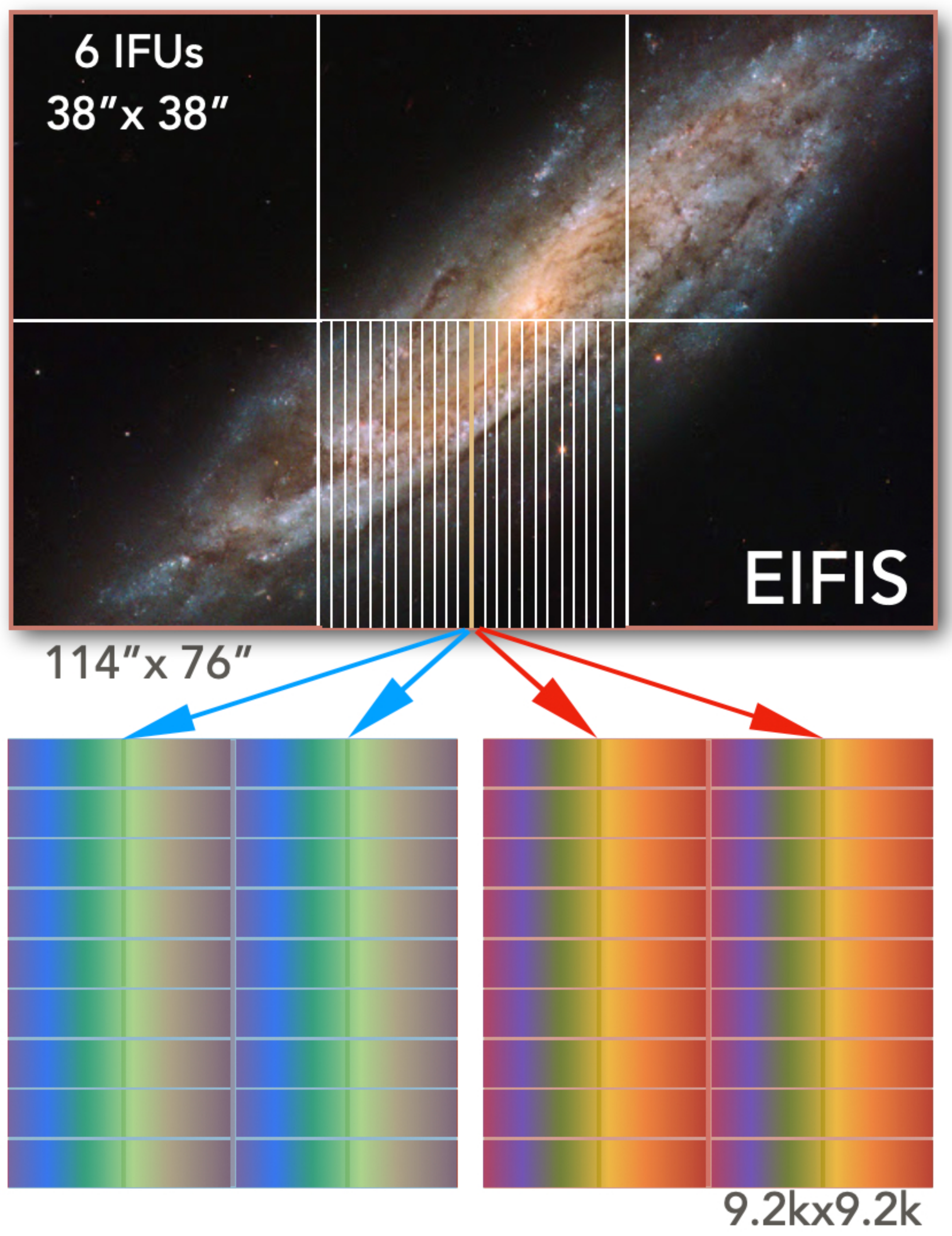}
	\end{center}
   \caption[] 
   { \label{fig:diagram} 
Schematic view of the setup on-sky of a six-spectrograph EIFIS indicating how the image slicer divides the light into the two (blue and red) arms and how two pseudoslits are projected onto the detector to double the field of view.}
   \end{figure} 

To maximise the spectral coverage, our proposed design splits the light in a blue and red arm using dichroics. The two spectrographs are fed light from the image slicer with the same pseudo-slits and share part of the collimating optics. With such a design one can cover the spectral range between 3000 and 5600 {\AA} in the blue arm and 5400 and 10100 {\AA} in the red arm at a resolving power of 2400. The small overlap is needed to cross-calibrate the two parts of the spectrum.

The proposed design is modular and hence can be adapted to use a different number of spectrographs and arrangements, resulting in a different sky area coverage. Our baseline design includes six  spectrographs to achieve a total field of view of 2.4 square arcminutes. In Fig.~\ref{fig:diagram} we present a diagram indicating how the sky area is split into the different spectrographs by the image slicers, reprojected and dispersed on the detectors of the two different spectral arms of each spectrograph. To test the feasibility of this design we conducted a first study to identify its limitations and search for the optimal design alternative.


\begin{table}[ht]
\caption{EIFIS specifications, comparing the proposed, optimal design and the result of a feasible design after the optomechanical study} 
\label{tab:Specs}
\begin{center}       
\begin{tabular}{|l|l|l|}
\hline
\rule[-1ex]{0pt}{3.5ex}  Item & Goal & Feasibility  \\
\hline
\hline
\rule[-1ex]{0pt}{3.5ex}  Individual FoV ($^{\prime\prime}$) & $38\times38$ & $29\times29$ \\
\hline
\rule[-1ex]{0pt}{3.5ex}  Number of Spectrographs & 6 & 8  \\
\hline
\rule[-1ex]{0pt}{3.5ex}  Total FoV (sqrarcmin) & 2.4 & 1.9  \\
\hline
\rule[-1ex]{0pt}{3.5ex}  Spectral range ({\AA}) & 3000-10100 & 3300-10100  \\
\hline
\rule[-1ex]{0pt}{3.5ex}  Resolving power ($\lambda/\delta\lambda$) & 2400 & 2400  \\
\hline
\rule[-1ex]{0pt}{3.5ex}  Spaxel size ($^{\prime\prime}$) & 0.3 & 0.4  \\
\hline 
\end{tabular}
\end{center}
\end{table}

\section{OPTICAL DESIGN}

Here we evaluate the feasibility of the optical design of EIFIS. Our analysis does not include the derotator that feeds the instrument, nor the image slicer that reprojects the sky area into a pseudo-slit at the entrance of each spectrograph (in our case two pseudo-slits), but only the optics after the  pseudo-slit. In search for the best possible option, we tested three different optical designs: A reflective design, a refractive one, and a combined one. Both the fully refractive and the combined design have issues with image quality and require a large number of optical elements, which could compromise the manufacturability and limit the efficiency of the instrument. The fully refractive design, however, would allow for a more compact design and hence to place a larger number of the modular spectrographs in the Nashmyth platform if required. The reflective design shows good optical results while being easier to manufacture and having higher efficiency, hence we adopt the reflective design as our baseline.

\begin{figure} [ht]
   \begin{center}
   \includegraphics[width=9cm]{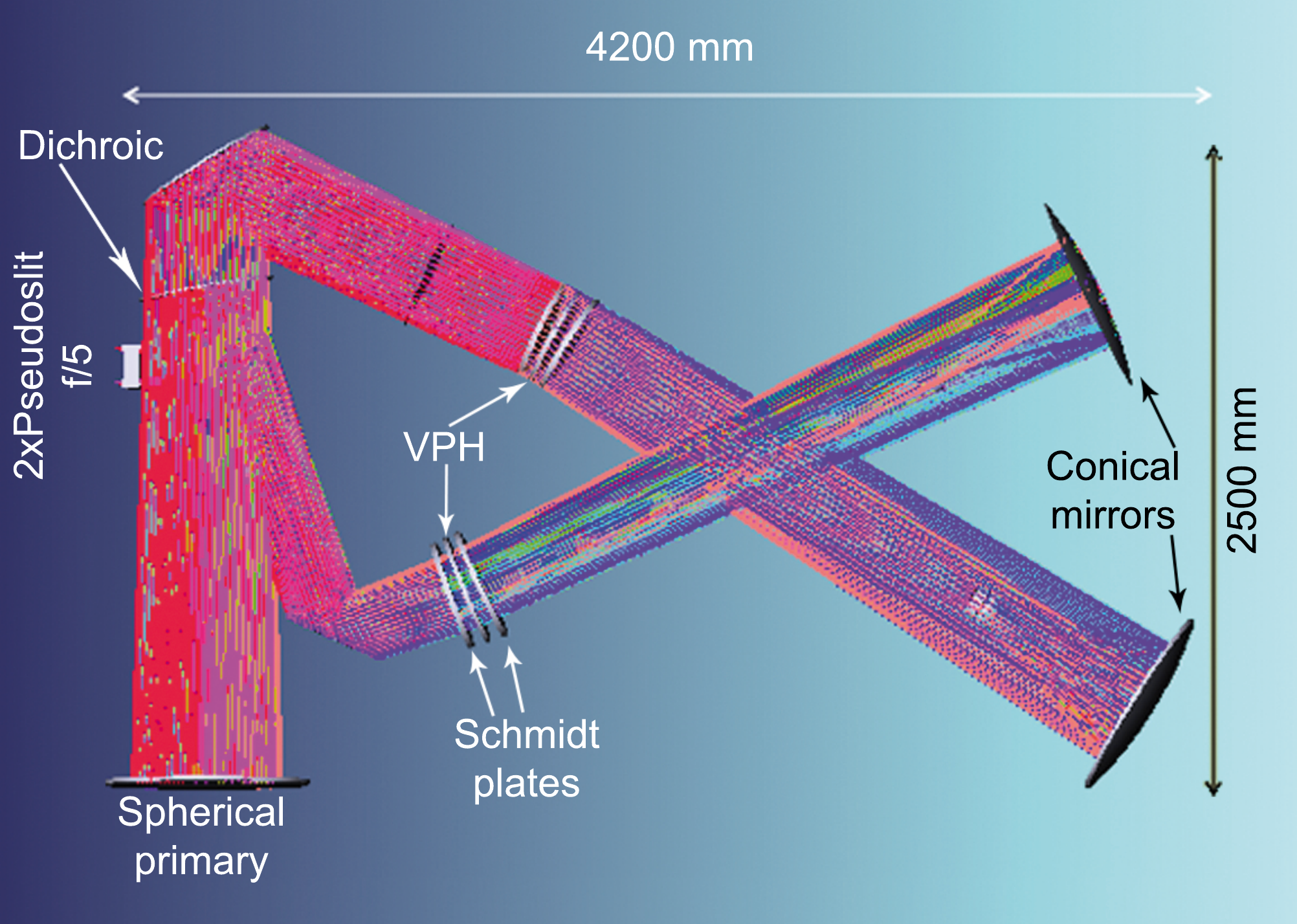}
	\end{center}
   \caption[] 
   { \label{fig:optics} 
Optical design of a double spectrograph using reflective optics for both collimator and camera.}
   \end{figure}

In the reflective design, we use two Schmidt camera optics, an inverted design for the collimator and a regular Schmidt design for the camera (see Fig. \ref{fig:optics}). The image slicer delivers an f/5 input, which is then directed to the first spherical mirror and gets reflected back onto a beam splitter, which separates the light into the blue and the red arm. Then the light in each arm goes through a corrector plate and gets dispersed with a volume phase holographic gratings (VPH) placed at the pupil. We assume a maximum pupil diameter of 400 mm to avoid too large physical sizes for the VPHs. Both red and blue arms hence use the same collimator mirror but different corrector plates. 

After the dispersion, the light enters the camera, another Schmidt design with a corrector plate immediately after the VPH. The initial design of the camera has an f-ratio of 2.0. The detector is then placed at the prime focus after a set of corrector lenses to flatten the focal plane. 

Each double spectrograph includes three curved mirrors, one spherical and two conical, the largest of which has a physical diameter of 800 mm. A flat mirror inserts the light from the slicer into the optical path of the camera, and a flat mirror in each arm is used to bend the light path and hence minimise the volume of the instrument. Further optics include one dichroic, two correctors (with two lenses each) and four field flatteners (with two lenses each). The VPHs at the pupil have a diameter of 400 mm to guarantee their manufacturability.

\begin{figure} [ht]
   \begin{center}
   \includegraphics[width=7.1cm]{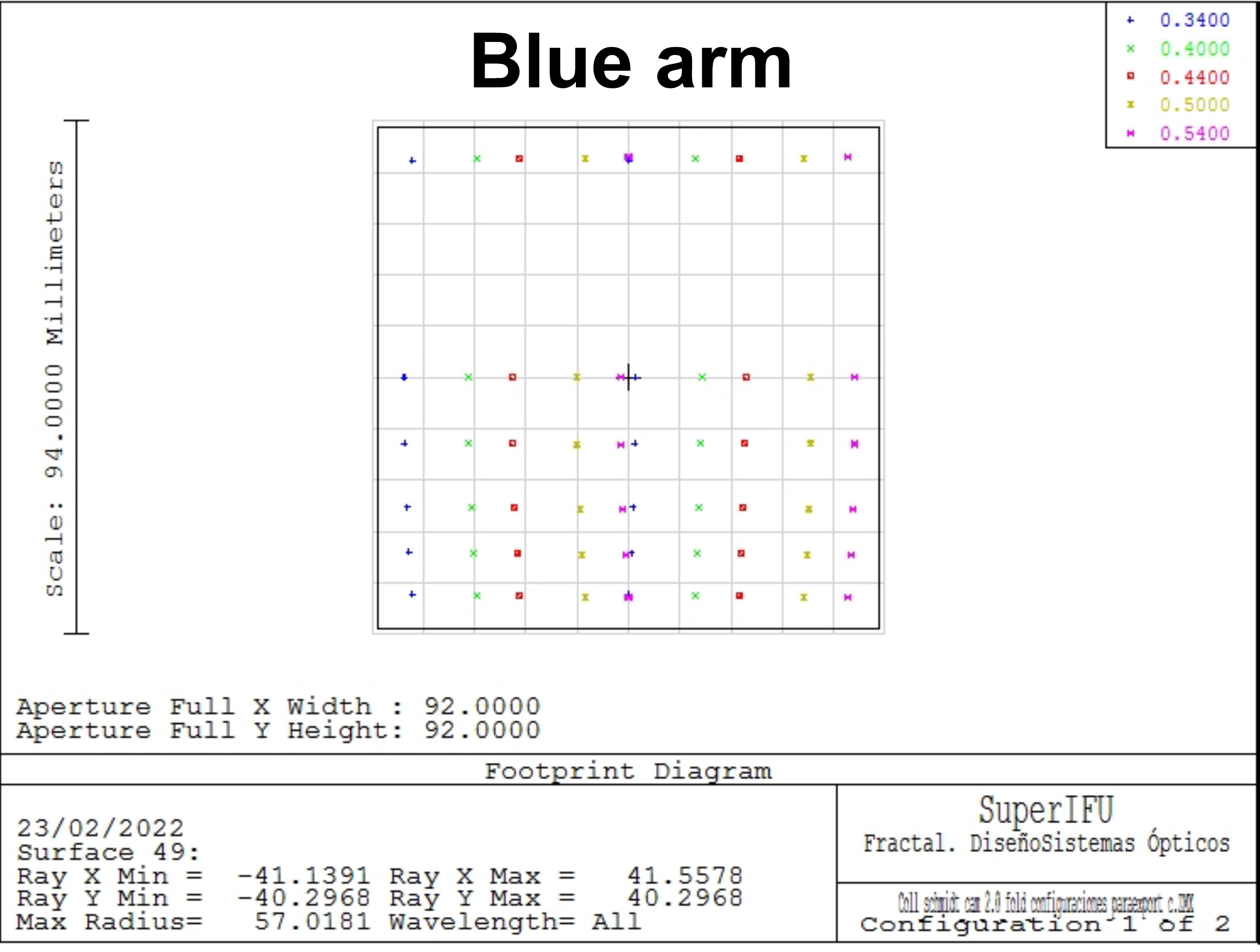}\includegraphics[width=7.1cm]{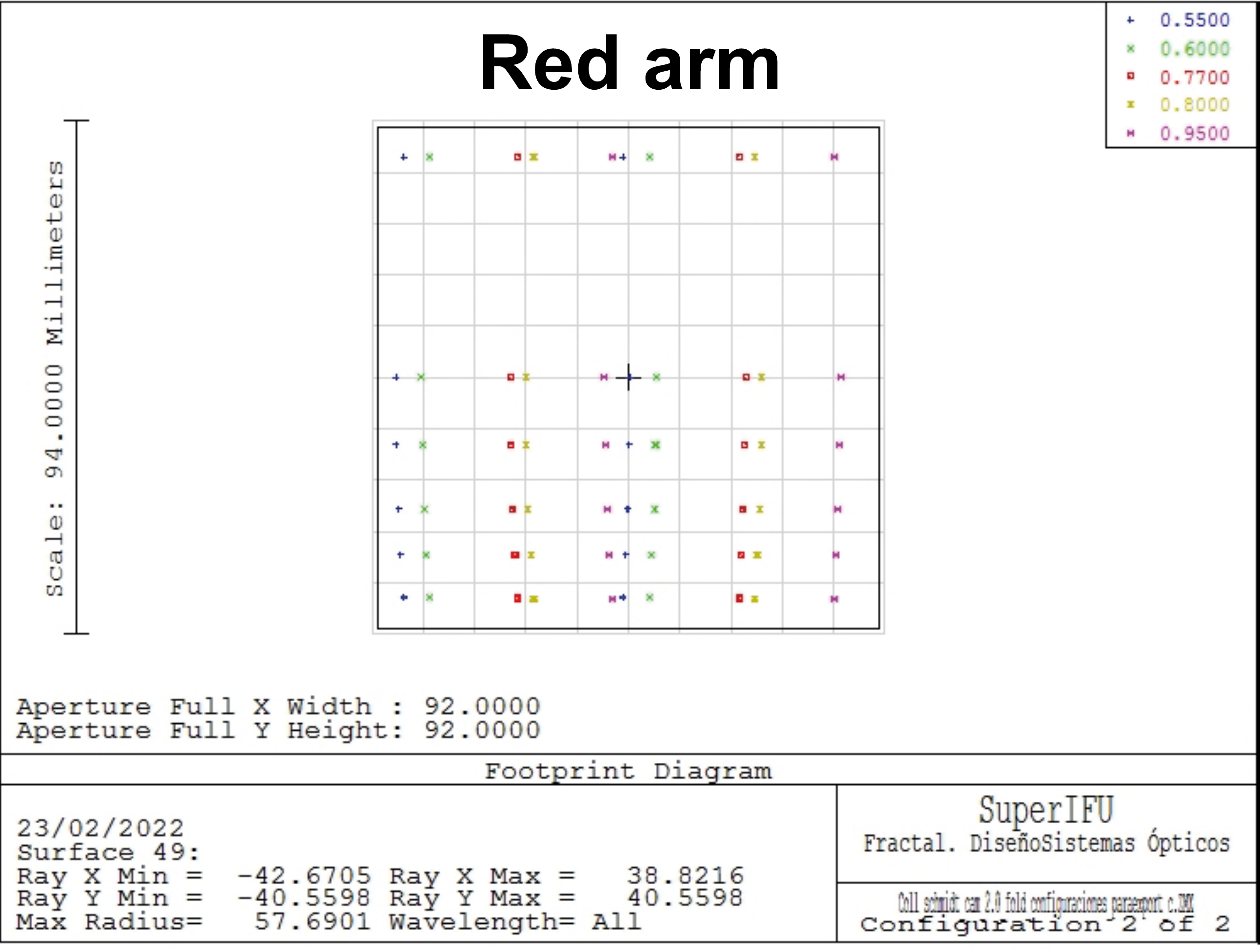}
      \includegraphics[width=14.2cm]{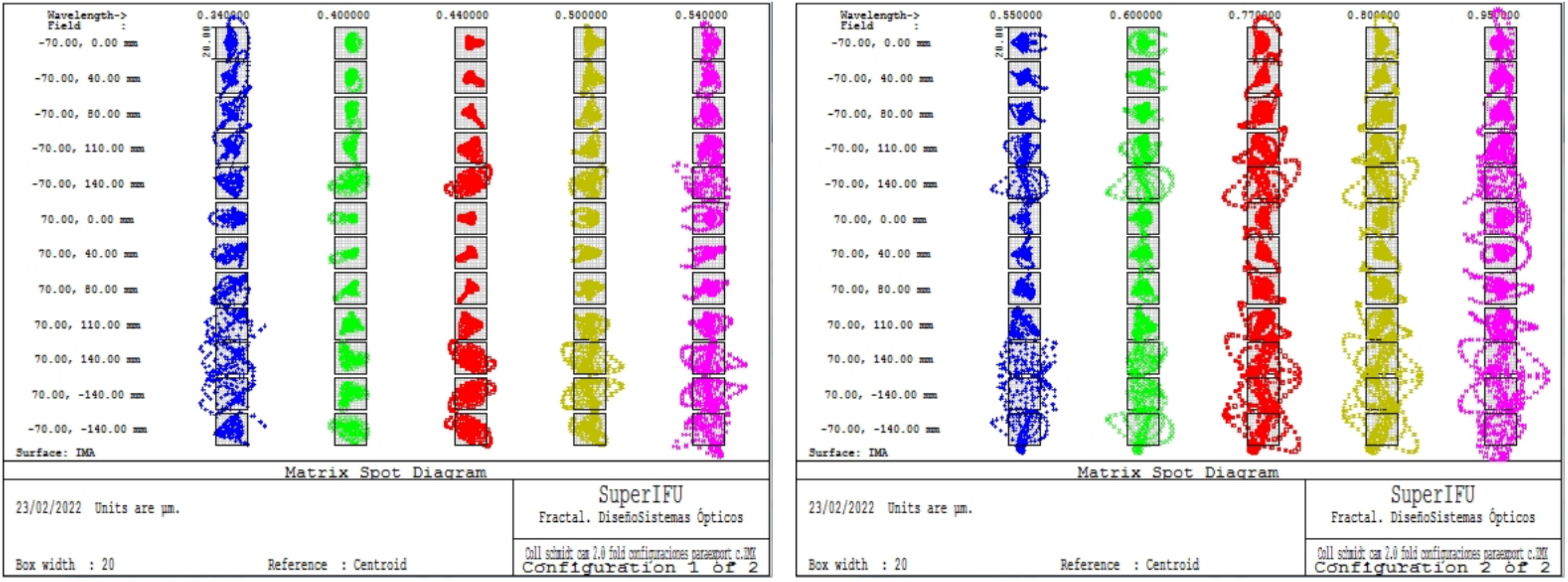}
	\end{center}
   \caption[] 
   { \label{fig:zemax} 
Top: Footprints of the spectra in the blue and red detectors, after the light is dispersed using a 250 lines per millimetre and 120 lines per millimetre VPH, respectively. Bottom: Spot diagrams for the red and blue arms, respectively.}
   \end{figure}

Figure~\ref{fig:zemax} shows the footprints and spot diagrams of both blue and red spectrograph arms. A preliminary analysis results in a very good image quality using a fully reflective design and the f/2 camera, and delivers a spot within 2 pixels for any location on the detector. An f/1.8 camera could be feasible at a moderate design effort.

\subsection*{TRADE-OFFS}

Our feasibility study shows that not all high-level requirements of the design can be simultaneously fulfilled, however, the viable options would still guarantee a very attractive instrument. In Table~\ref{table:tradeoffs} we present the different design options considering different spaxel sizes, field of view sizes, spectral coverage and resolving power. We select design 5 as a the best compromise between these quantities. The FOV is 20\% smaller and the spaxel size 30\% larger than the initial requirements. The spectral coverage is missing 300\AA{} at the blue end, where the efficiency would we rather low in any case. Design option 5 requires an f-ratio of 1.8, which would need some adjustments in the design. An f/2 camera design has been obtained without major refinement, hence we are confident that we can obtain good optical quality in a slightly faster camera as well.

\begin{table}[ht]
\caption{Trade-offs for the optical design. } 
\label{table:tradeoffs}
\begin{center}       
\begin{small}
\begin{tabular}{|l|cccc||c||c|}
\hline
\rule[-1ex]{0pt}{3.5ex}\textbf{Design}              & 1  & 2  & 3  & 4  & \textbf{*5*}  & 6  \\
\hline
\rule[-1ex]{0pt}{3.5ex}\textbf{Optical choices}     &   &   &   &   &   &   \\
\rule[-1ex]{0pt}{3.5ex}Camera focal ratio      & 2   & 2   & 2   & 1.8 & \textbf{1.8} & 1.8 \\
\rule[-1ex]{0pt}{3.5ex}Spaxel size ($^{\prime\prime}$)  & 0.3 & 0.4 & 0.5 & 0.3 & \textbf{0.4} & 0.5 \\
\rule[-1ex]{0pt}{3.5ex}Blue arm range ({\AA})& 3000--5600& 3500--5800& 3500--5600& 3300--5800& \textbf{3300--5800}& 3500--5700 \\
\rule[-1ex]{0pt}{3.5ex}Red arm range ({\AA}) &5600--10100&5800--9600&5600--9000&5800--10100&\textbf{5800--10100}&5700--9200 \\
\hline
\rule[-1ex]{0pt}{3.5ex}\textbf{Field of view }      &   &   &   &   &   &   \\
\rule[-1ex]{0pt}{3.5ex}Plate scale ($^{\prime\prime}$/pixel)& 0.103 & 0.103 & 0.103 & 0.115 & \textbf{0.115} & 0.115 \\
\rule[-1ex]{0pt}{3.5ex}FoV square side ($^{\prime\prime}$/pixel)&23.9&27.6&30.8&25.2&\textbf{29.1}&32.5\\
\rule[-1ex]{0pt}{3.5ex}Number of spectrographs & 8 & 8 & 8 & 8 & \textbf{8} & 8 \\
\rule[-1ex]{0pt}{3.5ex}Total FoV (arcmin$^2$) & 1.3 & 1.7 & 2.1 & 1.4 & \textbf{1.9} & 2.3 \\
\hline
\rule[-1ex]{0pt}{3.5ex}\textbf{Spectral resolution} &   &   &   &   &   &   \\
\rule[-1ex]{0pt}{3.5ex}Spaxel on detector (pixel) & 2.9 & 3.9 & 4.8 & 2.6 & \textbf{3.5} & 4.4 \\
\rule[-1ex]{0pt}{3.5ex}No. of resolution elements & 1584 & 1188 & 950 & 1760 & \textbf{1320} & 1056 \\
\rule[-1ex]{0pt}{3.5ex}Blue avg. disp. ({\AA}/elem) & 1.6  & 1.9  & 2.2  & 1.4  &\textbf{ 1.9 } & 2.1  \\
\rule[-1ex]{0pt}{3.5ex}Red avg. disp. ({\AA}/elem) & 2.8  & 3.2  & 3.6  & 2.4  & \textbf{3.3}  & 3.3  \\
\rule[-1ex]{0pt}{3.5ex}Blue avg. resolving power & 2620 & 2402 & 2059 & 3203 & \textbf{2403 }& 2208 \\
\rule[-1ex]{0pt}{3.5ex}Red avg. resolving power & 2763 & 2407 & 2041 & 3254 & \textbf{2441} & 2248 \\
\hline 
\end{tabular}
\end{small}
\end{center}
\end{table}

\section{MECHANICAL DESIGN}

\begin{figure} [!ht]
   \begin{center}
   \includegraphics[width=14cm]{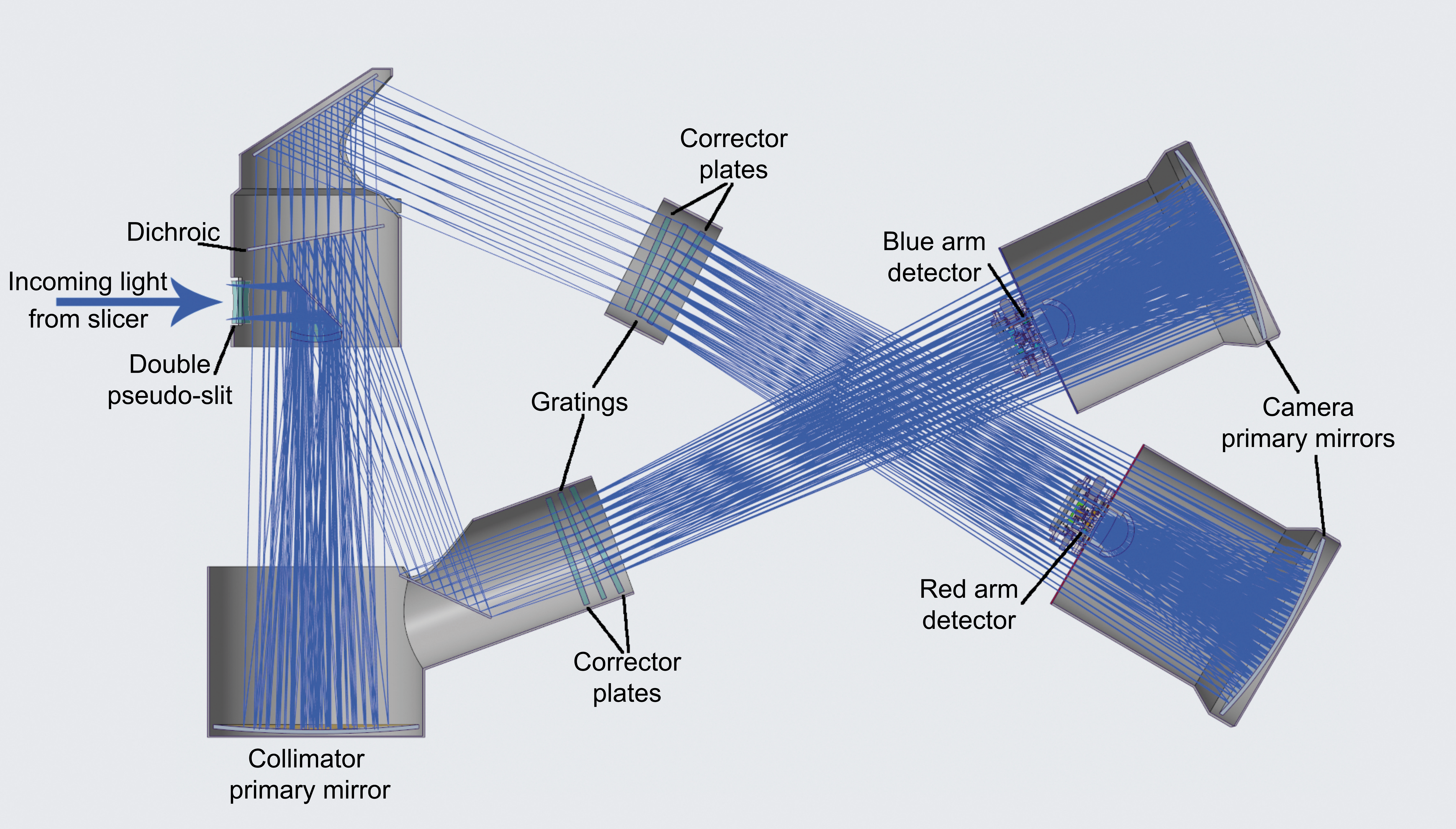}
	\end{center}
   \caption[] 
   { \label{fig:optomech} 
Opto-mechanical design of one of the double spectrographs}
   \end{figure}

\begin{figure} [!ht]
   \begin{center}
   \includegraphics[height=6cm]{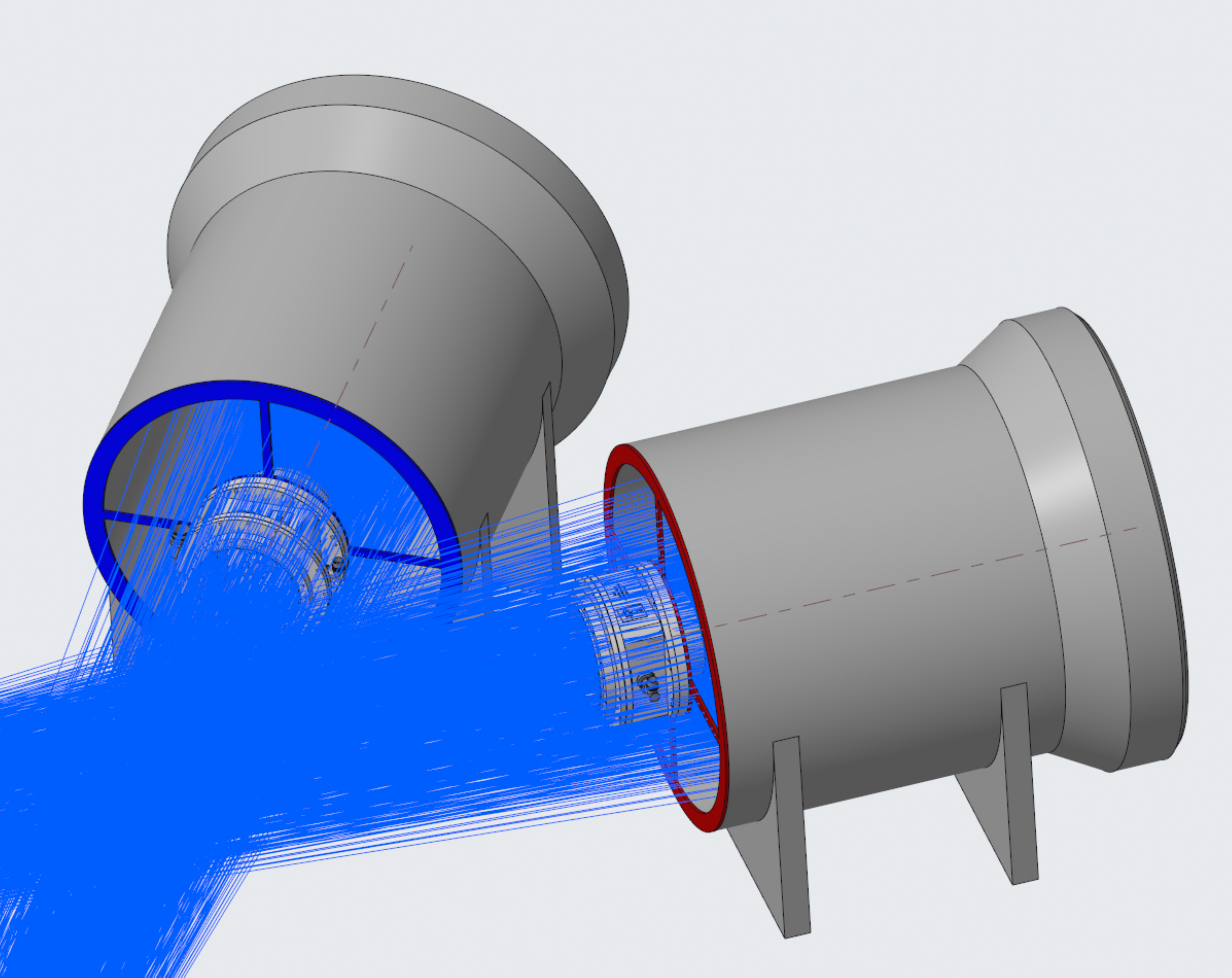}\includegraphics[height=6cm]{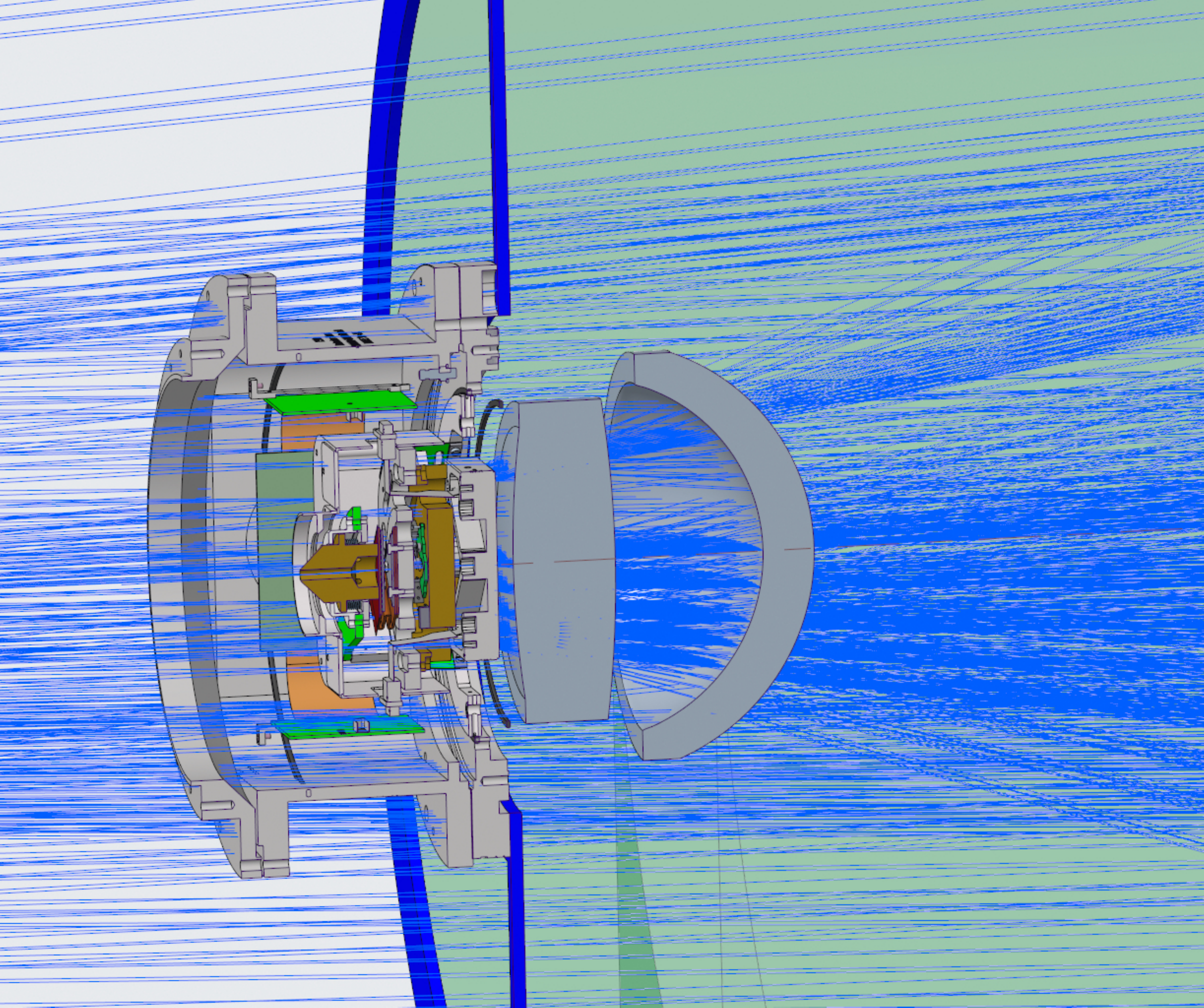}
	\end{center}
   \caption[] 
   { \label{fig:camera} 
Left: Detail of the opto-mechanical design of the cameras. Right: Detector placement within the focus of the camera.}
   \end{figure}

In this section we present a possible implementation of the mechanical design to evaluate the volume of each double spectrograph and attempt to place them in the Nashmyth platform. The optical elements have been grouped in several cylindrical enclosures to minimise differential rotations and to provide baffling. Further baffles would be included to eliminate stray light (see  Fig.~\ref{fig:optomech}). The detectors are located in the prime focus of the camera mirror. The camera casing and corrector lenses will be supported by a spider-like structure to minimise obstruction as shown in Fig.~\ref{fig:camera}.

\begin{figure} [!ht]
   \begin{center}
   \includegraphics[width=17cm]{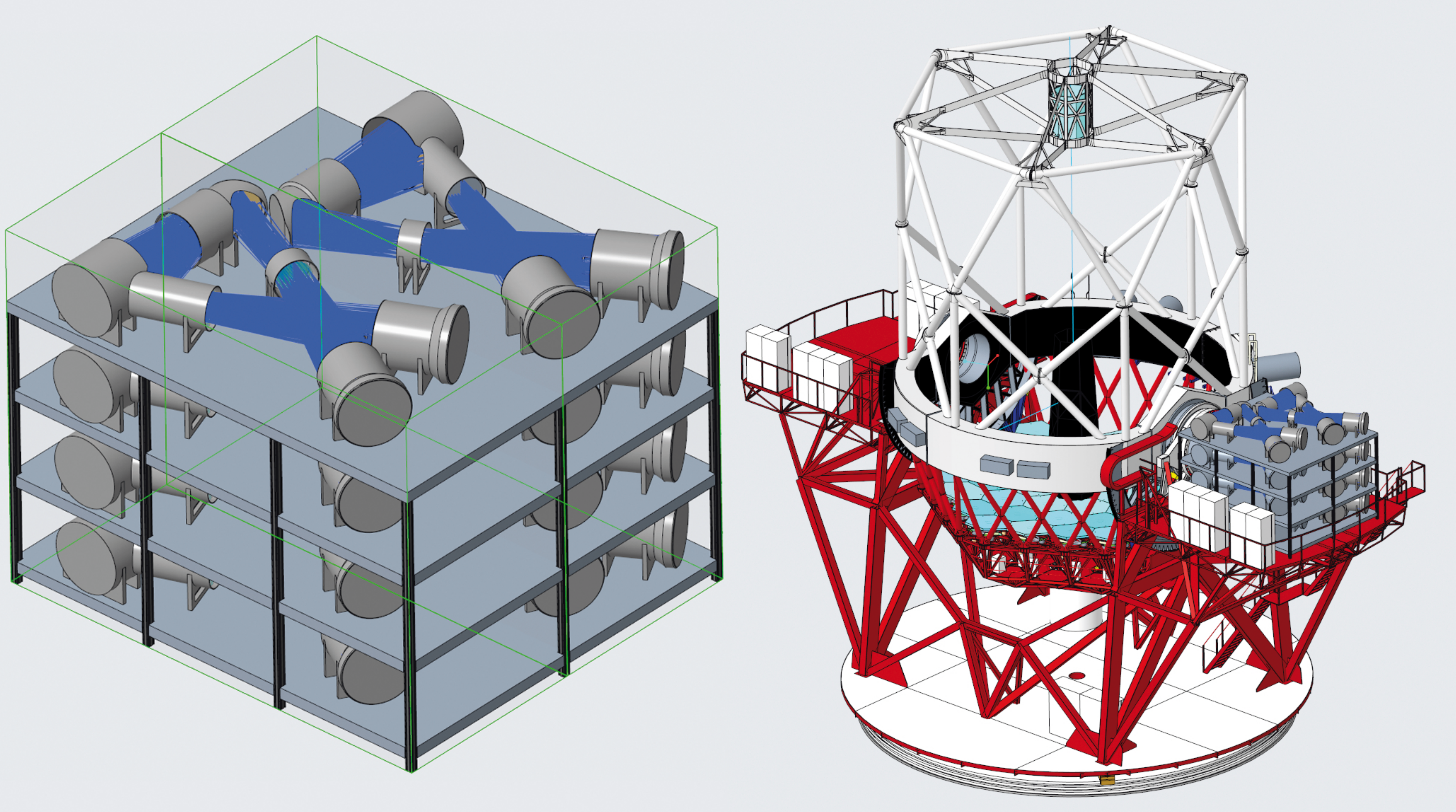}
	\end{center}
   \caption[] 
   { \label{fig:telescope} 
Envelope and placement for EIFIS with 8 double spectrographs filling the Nasmyth platform of the 10.4m GTC telescope. Left: The assembly of 8 spectrographs fits within the allocated space for instruments at the platform. Right: Rendering of the full telescope with the instrument on the platform.}
   \end{figure}

The compact arrangement of the spectrographs allows them to be piled on a  mechanical support structure in four rows such that up to 8 double spectrographs can be placed within the allocated space in the Nashmyth platform of GTC (see Fig.~\ref{fig:telescope}).

\section{CONCLUSIONS}

We conducted a feasibility study for a wide field spectrograph with full optical spectrum coverage for an 8-10 m class telescope, in this case aimed for the 10.4 m GTC. The design is based on a reflective double Schmidt camera design and uses two pseudo-slits to be imaged on a 9.2kx9.2k detector, which allows for a full wavelength coverage from 3000 - 10 000 \AA{}. To optimize the transmission and efficiency, the spectral range is split in two arms. To cover a large FOV a modular design has been chosen, which can be adapted to the scientific needs, budget and telescope. Although the initial specifications will be difficult to achieve, we have proven that we can construct an instrument with only small trade-offs in spectra resolution or FOV. Improvements have to be made concerning the mass and volume budget to fit the requirements of the GTC Nasmyth focus. We are currently waiting for a new instrument call at GTC to propose the design of EIFIS, alternatively, any other 8-10m class telescope could also be an option. The preferred placement would be in the Northern Hemisphere to complement MUSE and the future BlueMUSE instrument at VLT.

\acknowledgments 
CCT acknowledges partial funding from ASU-CAS to attend this conference. 
AdUP acknowledges financial support from the Côte D'Azur University through a \textit{CSI recherche} grant awarded to the GRANDMA project. JFAF acknowledges support from the Spanish Ministerio de Ciencia, Innovaci\'on y Universidades through grant PRE2018-086507 as well as support from Spanish National Research Project RTI2018-098104-J-I00 (GRBPhot).

\bibliography{report} 
\bibliographystyle{spiebib} 

\end{document}